\documentclass[aps,prl,twocolumn,superscriptaddress,showpacs]{revtex4}

\usepackage{graphicx}
\usepackage{latexsym}
\usepackage{amssymb}
\usepackage{amsmath}

\begin{document}

%\preprint{}

\title{Thermopower of Interacting GaAs Bilayer Hole Systems in the Reentrant Insulating Phase near $\nu=1$}

\author{S. Faniel}
\email[]{faniel@pcpm.ucl.ac.be}
\affiliation{Cermin, PCPM and DICE Labs, Universit\'e Catholique de Louvain, 1348 Louvain-la-Neuve, Belgium}

\author{E. Tutuc}
\affiliation{Department of Electrical Engineering, Princeton University, Princeton, New Jersey 08544, USA}

\author{E.P. De Poortere}
\affiliation{Department of Electrical Engineering, Princeton University, Princeton, New Jersey 08544, USA}

\author{C. Gustin}
\affiliation{Cermin, PCPM and DICE Labs, Universit\'e Catholique de Louvain, 1348 Louvain-la-Neuve, Belgium}

\author{A. Vlad}
\affiliation{Cermin, PCPM and DICE Labs, Universit\'e Catholique de Louvain, 1348 Louvain-la-Neuve, Belgium}

\author{S. Melinte}
\affiliation{Cermin, PCPM and DICE Labs, Universit\'e Catholique de Louvain, 1348 Louvain-la-Neuve, Belgium}

\author{M. Shayegan}
\affiliation{Department of Electrical Engineering, Princeton University, Princeton, New Jersey 08544, USA}

\author{V. Bayot}
\affiliation{Cermin, PCPM and DICE Labs, Universit\'e Catholique de Louvain, 1348 Louvain-la-Neuve, Belgium}

\date{\today}

\begin{abstract}
We report thermopower measurements of interacting GaAs bilayer hole systems. When the carrier densities in the two layers are equal, these systems exhibit a reentrant insulating phase near the quantum Hall state at total filling factor $\nu=1$. Our data show that as the temperature is decreased, the thermopower diverges in the insulating phase. This behavior indicates the opening of an energy gap at low temperature, consistent with the formation of a pinned Wigner solid. We extract an energy gap and a Wigner solid melting phase diagram.
\end{abstract}


\maketitle

The ground state of two-dimensional (2D) carrier systems has been subject to intensive work over the past decades. The interplay of Coulomb interaction, disorder, and magnetic field $B$ leads to many different ground states~\cite{DoubleQHEReview}. In low-disorder 2D systems, a sufficiently large $B$ is expected to quench the kinetic energy of the carriers, leading to the formation of a Wigner solid (WS)~\cite{WC}. Experimental evidence has been accumulated for the existence of  a WS in both GaAs 2D electron single-layers~\cite{DoubleQHEReview,WC,Jiang90,Goldman} around the fractional quantum Hall state at Landau level filling factor $\nu =$ $\frac{1}{5}$ and electron bilayers~\cite{Manoharan96} in the vicinity of total filling factor $\nu =$ $\frac{1}{2}$, as well as in dilute, single-layer, GaAs 2D hole systems~\cite{Santos92} near $\nu =$ $\frac{1}{3}$. 

Recently, a reentrant insulating phase (IP) was observed in interacting GaAs {\em bilayer} hole systems around the many-body quantum Hall state (QHS) at total filling factor $\nu=1$~\cite{Tutuc03}. It was suggested in Ref.~\cite{Tutuc03} that this IP signals a bilayer WS, stabilized at such high $\nu$ by strong Landau level mixing due to higher effective mass of holes, and by the interlayer Coulomb interaction in the bilayer system. While magnetotransport studies in bilayer systems have intensified in recent years~\cite{Kellogg03}, little attention was paid to their thermoelectric properties. Thermopower coefficients are sensitive tools to study the electronic properties of 2D systems, including Si inversion layers~\cite{ThermopowerSi}, SiGe quantum wells~\cite{ThermopowerSiGe}, and GaAs heterojunctions~\cite{ThermopowerGaAs1,ThermopowerGaAs2,Bayot94}. In particular, as the temperature $T \rightarrow 0$, the diffusion thermopower is expected to vanish as $T^{1/3}$ for a mobility gap or diverge as $T^{-1}$ in the presence of an energy gap~\cite{Chaikin}.

We study here the thermopower of interacting GaAs bilayer hole systems around $\nu=1$. When the hole densities in the two layers are equal, at filling factors near $\nu=1$, our measurements  show that  the diffusion thermopower diverges as  $T^{-1}$  when the temperature is lowered below  $\sim$ 100 mK. This behavior corroborates the magnetotransport~\cite{Tutuc03} and the non-linear $I-V$ results. The $T$ dependence of the thermopower  indicates the opening of an energy gap, consistent with the formation of a pinned, bilayer WS near the QHS at $\nu=1$, and allows us to extract a phase diagram for the WS melting.

The samples we have studied are Si modulation-doped GaAs bilayer hole systems grown by molecular beam epitaxy on undoped GaAs (311)A wafers. The two samples consist of two 15 nm wide GaAs quantum wells separated by an AlAs barrier with a thickness of 11 nm (sample A) and 7.5 nm (sample B). The width of the AlAs barrier and the large effective mass of the holes assure a weak interlayer tunneling~\cite{DSAS}.  Two $8 \times 2 \rm \ mm^{2}$ rectangular pieces of wafer were cut with their length along the [$01\bar{1}$] (sample A) and [$\bar{2}33$] (sample B) crystallographic directions. On both specimens, we patterned $400 \ \rm \mu m$ wide Hall bars in order to reduce density inhomogeneities; the distance between the voltage probes along the Hall bar is 1.2 mm. Ohmic contacts were made to both layers. A back-gate was used  to tune the charge distribution of the two layers. The charge imbalance is denoted by $\Delta p = (p_T-p_B)/2$, where $p_B$ and $p_T$ are the hole densities of the bottom and the top layer, respectively. For balanced charge distributions ($\Delta p = 0$), the total hole density for both samples is close to $p = 5 \times  10^{11} \ \rm{cm^{-2}}$~\cite{Footnote2}. Our experimental setup for thermopower measurements is described in Ref.~\cite{Bayot94}. One end of each sample was mounted onto the cold finger of a dilution refrigerator, while a strain gauge, used as heater, was glued on the other end using GE varnish. To measure the longitudinal thermopower $S_{xx}$, a thermal gradient was created along the sample by applying a sine wave current at a frequency $f=1.3$ Hz through the heater and measuring, via a lock-in amplifier, the voltage signal at $2f$ between two contacts on the same edge of the Hall bar. The temperature gradient was determined using two carbon paint thermometers deposited on the sample's surface, and was kept below 10$\%$ of the mean $T$.

\begin{figure}[htbp]
  \includegraphics[width=1\columnwidth]{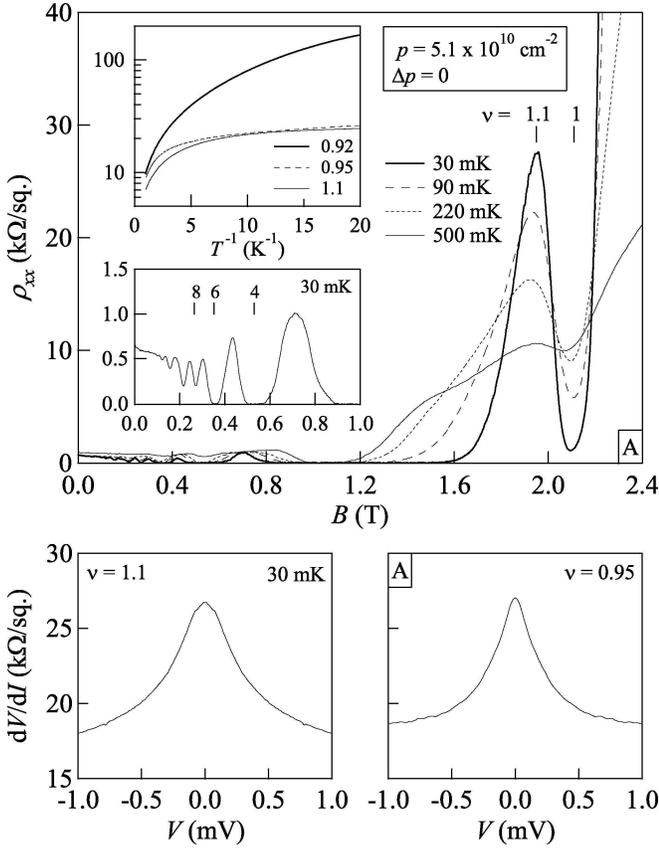}
  \caption{\label{Rxx}Upper panel: longitudinal resistivity $\rho_{xx}$ as a function of $B$ at the indicated temperatures for sample A at a total density $p = 5.1 \times  10^{11} \ \rm{cm^{-2}}$. For the data in the figure, the densities of the individual layers are equal ($\Delta p = 0$). Top inset: The $T$ dependence of $\rho_{xx}$ at  $\nu=0.92$, 0.95, and 1.1. Bottom inset: A magnified view of the low-$B$ $\rho_{xx}$ trace at 30 mK. Lower panels: differential resistance in the insulating phase of sample A as a function of the dc bias for $\nu=1.1$ (left panel) and 0.95 (right panel) at 30 mK.}
\end{figure}

We first summarize the magnetotransport results for our samples. In Fig.~\ref{Rxx}, we illustrate for sample A the longitudinal resistivity $\rho_{xx}$ as a function of $B$ at low $T$. Consistent with the sample's extremely small interlayer tunneling~\cite{DSAS}, at low B the $\rho_{xx}$ trace exhibits minima only at even integer fillings $\nu = 2$, 4, 6, etc. The strong interlayer interaction, however, leads to a many-body QHS at $\nu = 1$~\cite{Kellogg03}, flanked by reentrant IP's~\cite{Tutuc03}. The measured QHS excitation gaps $^1\Delta$, extracted from the $T$ dependence of $\rho_{xx}$ at $\nu = 1$, are 0.13 and 0.67 K, respectively; the larger $^1\Delta$ for sample B is consistent with its smaller $d/l_B$ ratio~\cite{Footnote2}. The $T$ dependence of $\rho_{xx}$ at  $\nu=0.92$, 0.95, and 1.1 is displayed in the top inset to Fig.~\ref{Rxx}, revealing the insulating behavior at these fillings. Finally, in the bottom panels of Fig.~\ref{Rxx} we show $I - V$ characteristics at filling factors near $\nu=1$ for sample A, exhibiting a strongly non-linear behavior. We have observed qualitatively similar $I - V$ curves for sample B also. These non-linear data are similar to previous measurements in the IP near $\nu=\frac{1}{3}$ in single-layer GaAs 2D hole systems~\cite{Santos92}.

Before discussing our $S_{xx}$ data, we briefly review the essential concepts for understanding thermopower data in 2D carrier systems~\cite{TP2DEG}. In general $S_{xx}$ consists of two additive contributions: the phonon-drag thermopower $S^g_{xx}$ and the diffusion thermopower $S^d_{xx}$. In GaAs-based heterostructures, the former is known to be dominant for $T$  $\gtrsim 200 \ \rm mK$, while the latter provides the main contribution at lower temperatures ($T$  $\lesssim 100 \ \rm mK$). 
The phonon-drag thermopower is given by $S^g_{xx} = \frac{C}{3pe} \frac{\Lambda}{v_s \tau_{ph}}$, where $C$ is the specific heat, $\Lambda$ is the phonon mean free path, $e$ is the electron charge, $v_s$ is the sound velocity, and $\tau_{ph}$ is the phonon momentum relaxation time due to scattering with holes only~\cite{TP2DEG}.
The diffusion thermopower, on the other hand, measures the heat or the entropy per carrier and is sensitive to the density of states at the Fermi level. In particular, for a 2D insulating system, $S^d_{xx}$ can distinguish between a mobility gap and an energy gap. In the first case, $S^d_{xx}$ should vanish at low $T$ as $S^d_{xx} \propto T^{1/3}$, while for the second case - when an energy gap $E_g$ separates the ground state from its excitations - $S^d_{xx}$ is expected to diverge at low $T$ as~\cite{Chaikin,Ziman,FootnoteReferee}: 

\begin{equation}
S^d_{xx} \cong \frac{k_B}{2e} \frac{E_g}{k_B} \ T^{-1}.
\label{Eq1}
\end{equation}

\noindent Hence, $S^d_{xx}$ is a unique probe of the ground state of 2D  systems as it can critically test if the IP results from the strong localization of the carriers (mobility gap) or from the formation of a pinned WS (energy gap).  

\begin{figure}[htbp]
  \includegraphics[width=1\columnwidth]{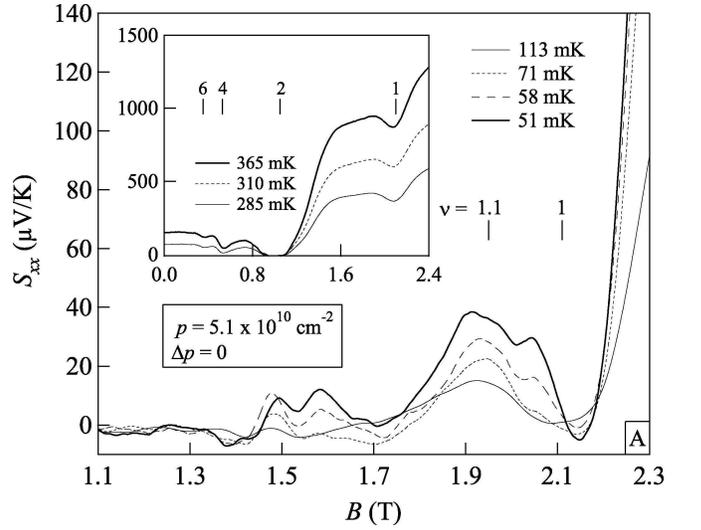}
  \caption{\label{MagnetoThermopower}$S_{xx}$ as a function of $B$ at four temperatures  below $120 \ \rm mK$ for sample A. Inset: $S_{xx}$ vs $B$ is displayed at $T = 285$, 310, and $365 \ \rm mK$ in the phonon-drag regime. }
\end{figure}

\begin{figure}[htbp]
  \includegraphics[width=1\columnwidth]{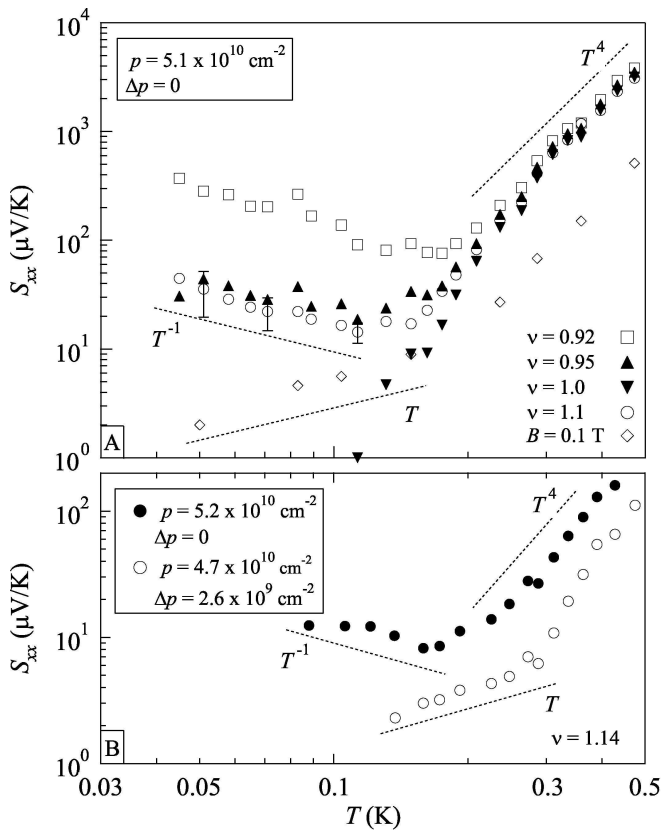}
  \caption{\label{TPvsT}Top panel (sample A): $T$ dependence of $S_{xx}$ at different filling factors when the bilayer is balanced. Bottom panel (sample B): $T$ dependence of $S_{xx}$ at $\nu$ = 1.14 for balanced ({\large $\bullet$}) and  imbalanced ({\large $\circ$}) charge distributions. The dotted lines indicate various $T$ dependencies. The error bars are given by the noise of the lock-in amplifier, the magnitude of the applied thermal gradient, and the accuracy of the thermometers' calibration.}
\end{figure}

We now present the result of our $S_{xx}$ measurements, summarized in Figs.~\ref{MagnetoThermopower} and~\ref{TPvsT}, and consider first the phonon-drag regime. Figure~\ref{MagnetoThermopower} shows $S_{xx}$ vs $B$ traces for sample A at various temperatures. The curves at $T = 285$, 310, and $365 \ \rm mK$, where the phonon-drag is dominant, are displayed in the Fig.~\ref{MagnetoThermopower} inset. Similar to $\rho_{xx}$ data, we observe minima at even filling factors ($\nu=2$, 4, and 6) and at $\nu=1$ consistent with a vanishing density of states at the Fermi level~\cite{TP2DEG}. In the phonon-drag regime, a dramatic decrease of $S_{xx}$ is found in the entire investigated $B$ range as $T$ diminishes. This is clearly indicated by
the $T$ evolution of $S_{xx}$ at various fillings shown in Fig.~\ref{TPvsT} for both samples.
In the top panel of Fig.~\ref{TPvsT}, $S_{xx}$ vs $T$ is displayed for sample A at $\nu$ = 1.1, 1, 0.95, 0.92 and for $B=0.1$ T when the bilayer system is balanced. In the bottom panel of the figure, we show $S_{xx}$ vs $T$ for sample B. Data points are presented at $\nu$ = 1.14 for a balanced and a slightly imbalanced charge distribution. At high $T$, $S_{xx} \cong S^g_{xx}$ follows the well known $T^4$ law~\cite{Lyo} for all the experimental configurations shown in Fig.~\ref{TPvsT}~\cite{PhononDrag}.

At low temperatures, where the thermopower is dominated by the diffusion contribution, $S_{xx}$ increases with decreasing $T$ in the IP's reentrant around $\nu = 1$. This main result of our work is illustrated in Fig.~\ref{MagnetoThermopower} by the $S_{xx}$ vs $B$ traces at $T =  113$, 71, 58, and 51 mK~\cite{Footnote3}. The divergence of $S_{xx}$ at low $T$ in the IP is further illustrated in Fig.~\ref{TPvsT}. Focusing on the data shown in the top panel of Fig.~\ref{TPvsT}, we observe that at $\nu=1$, $S_{xx}$ rapidly approaches zero as expected for a QHS with no entropy~\cite{Chaikin} and $S_{xx}$ at $B=0.1$ T is proportional to $T$ below $200 \ \rm mK$ as expected for a diffusion-dominated thermopower in a metallic system. However at $\nu=1.1$, 0.95, and 0.92, $S_{xx} \propto T^{-1}$ as $T \rightarrow 0$. This divergence of $S_{xx}$ in the reentrant IP's at very low $T$ indicates the opening of an energy gap and suggests the formation of a pinned bilayer WS near $\nu = 1$.

Data for sample B exhibit qualitatively the same behavior. In the lower panel of Fig.~\ref{TPvsT} we show $S_{xx}$ vs $T$ data for sample B at $\nu=1.14$ where this sample's magnetotransport data reveal an IP. The $S_{xx}$ data show a divergence at low $T$ when the bilayer system is balanced (closed circles, $\Delta p = 0$). In the same panel we also show data (open circles, $\Delta p = 2.6 \ \times \ 10^{9} \ \rm cm^{-2}$) for sample B when charge distribution is imbalanced, i.e., when charge is removed from the the bottom layer via the application of positive back-gate bias. For the imbalanced state, $S_{xx}$ at $\nu=1.14$ turns from a $T^{-1}$ to an approximatively linear temperature dependence. This observation is in agreement with magnetotransport results showing that the reentrant IP disappears for small charge imbalance between the two layers~\cite{Tutuc03}. We note that the strengthening of the $\nu=1$ QHS when $\Delta p$ increases, seen in the magnetotransport~\cite{Tutuc03}, is also observed in thermopower measurements.
In other words, the filling factor range around $\nu=1$ where $S_{xx} \rightarrow 0$ as $T \rightarrow 0$ extends when charge is transfered from one layer to the other (data not shown here). 

\begin{figure}[htbp]
  \includegraphics[width=1\columnwidth]{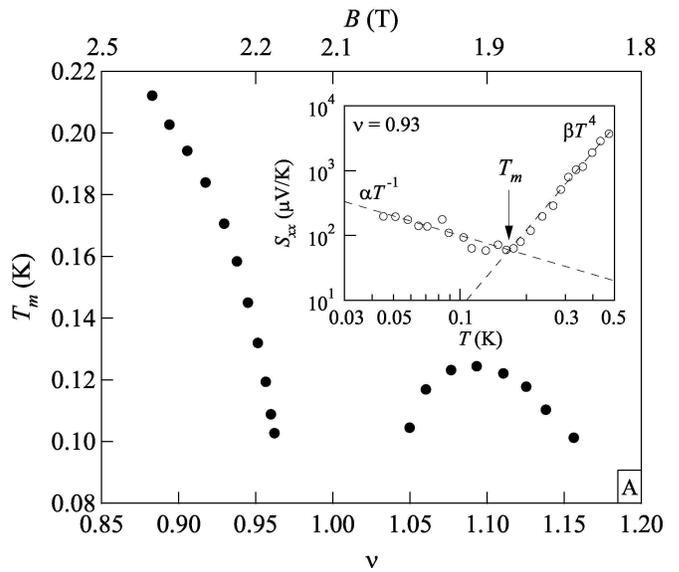}
  \caption{\label{TmVsB}Evolution of the melting temperature $T_m$ as a function of $\nu$ of the pinned, bilayer WS in sample A. The top axis shows the corresponding magnetic fields. Inset: $T_m$ is extracted from the $T$ dependence of $S_{xx}$ by assuming that the rise in $S_{xx}$ signals the melting of the WS.}
\end{figure}

We now turn to a more quantitative discussion of our thermopower data. From the $T$ dependence of $S_{xx}$, shown in Fig.~\ref{TPvsT} and the inset to Fig.~\ref{TmVsB}, we can extract a phase diagram for the WS melting. To estimate the melting temperature $T_m$, we assume that the solid-to-liquid transition occurs when $S_{xx}$ starts to rise as $T$ is lowered. As illustrated in the inset to Fig.~\ref{TmVsB} for $\nu=0.93$, we have fitted at fixed filling factors the phonon-drag thermopower with a $T^4$ law for $T \gtrsim 200 \ \rm mK$ and the diffusion thermopower with a $T^{-1}$ law for $T \lesssim 100 \ \rm mK$. The intersection of these two fits determines $T_m (\nu)$. In Fig.~\ref{TmVsB}, the $T_m$ vs $\nu$ phase diagram is shown for sample A. The reentrant behavior of the IP is illustrated by the evolution of $T_m$; $T_m$ is finite in a small filling range centered near $\nu=1.1$ and at $\nu \gtrsim 0.95$, but is vanishingly small near $\nu=1$ ($B=2.1$ T) where the quantum Hall ground state is present. The magnitude of the estimated $T_m$ in our sample is similar to the reported values of the WS melting temperature in single-layer 2D hole~\cite{Bayot94} and electron~\cite{Goldman,Paalanen} systems around $\nu =$ $\frac{1}{3}$ and $\frac{1}{5}$, respectively. Using Eq.~(\ref{Eq1}) and the $S_{xx} = \alpha T^{-1}$ fit to the diffusion regime, an estimate of the energy gap associated with the pinned WS can be made when $E_g \gg k_{\rm B}T$~\cite{Bayot94}. This energy gap likely corresponds to the creation energy of thermally activated defects (such as bound pairs of dislocations), which may be responsible for electrical conduction in the pinned WS~\cite{Chui}. For example, at $\nu=0.92$, $E_g/k_{\rm B}=0.34 \ \rm K$ in sample A. This value is of the same order of magnitude as the activation energy $E_A/k_{\rm B}=0.15 \ \rm K$ determined from fitting $\rho_{xx}$ in the low-$T$ range to $\ln (\rho_{xx}) \propto E_A/2k_{\rm B}T$, and compares well with the WS melting temperature $T_m = 0.18 \ \rm K$ at this filling. 

To conclude, we have measured the thermopower of interacting GaAs bilayer hole systems around $\nu=1$. At very low $T$, when the diffusion thermopower dominates, $S_{xx}$ diverges in the IP's reentrant around $\nu=1$. This behavior indicates the opening of an energy gap and the presence of a pinned bilayer WS. We have interpreted our thermopower data in terms of a WS melting and extracted a $T_m$ vs $\nu$ phase diagram.

\begin{acknowledgments}
The work was supported by the DOE and the NSF, the von Humboldt Foundation, "Actions de recherches concert\'ees (ARC) - Communaut\'e fran\c{c}aise de Belgique", and by the Belgian Science Policy through the Interuniversity Attraction Pole Program PAI (P5/1/1). S.F. acknowledges financial support from the F.R.I.A.
\end{acknowledgments}



\begin{thebibliography}{10}
\bibitem{Footnote2} The ratio of intralayer and interlayer energies, given by $d/l_B$ where $d$ is the interlayer distance and $l_B$ is the magnetic length at $\nu=1$, is 1.47 for sample A and 1.29 for sample B.
\bibitem{FootnoteReferee} It is useful to describe $S=j_Q/(j_ET)$, where $j_Q$ and $j_E$ are the microscopic heat and electric currents, respectively. If the drift velocity of charge carriers is $v_d$ and their mobility $\mu$, then $j_E=pev_d$, $j_Q=Qv_d$, and $\rho=1/(pe\mu)$. When the charge carriers are excited across an energy gap $E_g$, the heat current is $j_Q=\frac{E_g}{2}pv_d$ as the characteristic heat of carriers is the difference between the band edge energy and the chemical potential, i. e. $Q \cong \frac{E_g}{2}$. The thermopower is then $\cong (\frac{k_B}{2e})(\frac{E_g}{k_BT})$ and diverges as $T \rightarrow 0$. Therefore the microscopic reason of thermopower's divergence is that the carriers carry a minimum amount of heat $E_g/2$, while the divergence of the resistivity is due to the fact that less and less carriers are excited across the gap as $T \rightarrow 0$.
\bibitem{Footnote3}In the field range $1.3 < B < 1.7$ T, we observe a weak (non-zero), oscillating $S_{xx}$ signal. We do not know the origin of these features. We speculate that they may be signaling the presence of fractional QHS's (e.g. at $\nu=\frac{4}{3}$), and a competition between such states and the engulfing IP. We note that we observe some of these features (e.g., an $S_{xx}$ minimum near $\nu=\frac{4}{3}$) in some $S_{xx}$ vs $B$ traces of sample B also. Note that, interestingly, $\rho_{xx}$ in this field range is rather featureless and shows a deep minimum ($\rho_{xx}>0$).
\end{thebibliography}
\end{document}